# Generalization of Cramer's rule and its application to The projection of Hartree-Fock wave function

**M. Hage-Hassan**
Université Libanaise, Faculté des Sciences Section (1)
Hadath-Beyrouth

## Abstract

We generalize the Cramer's rule of linear algebra. We apply it to calculate the spectra of nucleus by applying Hill-Wheeler projection operator to Hartree-Fock wave function, and to derive Löwdin formula and Thouless theorem. We derive by an elementary method the infinitesimal or Löwdin projection operators and its integral representation to be useful for the projection of Slater determinant.

## 1. Introduction

The Hartree-Fock variationnelle method provides an approximate determination of ground states and ground state energies of quantum mechanical systems, and widely used in physics and chemistry. In Hartree-Fock method [1-3] we approximate the ground state of the system by a Slater determinant $|\Phi_{HF}\rangle$ constructed from the states of nucleons which are eigenstates of a single particle hamiltonien called Hartree-Fock hamiltonien. This wave function is not function of angular momentum, and the calculation of rotational energy [3, 5, 7] can be done by using the integral representation of Hill-Wheeler operator or using the infinitesimal projection operators [8]. The great recent interest [12-16] to study the projection theory and their application in nuclear physics leads me to resume my former works on the projection of angular momentum [11].
  Löwdin [3, 7] proposed a formula for the calculation of the spectrum of energy levels, but this method requires a long calculation [1, 9-12] and does not take account the conditions of stability resulting from the minimization of the energy of the system.
   We observe that the calculation of the rotational energy using the Hartree-Fock theory implies the calculation of a determinant, the overlap of rotation, and a set of determinants which differs from it by the change of two columns [2-3, 11-12]. This leads us to the generalization of Cramer's rule of linear algebra [11] allowing us to calculate all these determinants by Gauss Elimination method.
This method takes into account the conditions of stability and minimizes the time of executions. Using this generalization we derive the Löwdin formula and the well known Thouless theorem [17, 18].

The infinitesimal projection operators of Löwdin [19-21] was ignored or "not known to most physicists" [20], despite the important mathematical works that was devoted to extend it to the Lie groups. The calculations using this operator is not easy and especially to project a Slater determinant. To overcome this difficulty, we present an elementary and general method for the determination of infinitesimals projection operators. Thus, we find the projections operators for the harmonic oscillator, for angular momentum and their relationship with the Gauss function. Then we find the integral representation of Löwdin operator useful for the projection of Slater determinant.

We present in the second section the generalization of Cramer's rule. We present in the third section a brief review of Hartree-Fock and the projection methods of the wave function to obtain the spectra of nuclei. In the fourth section we apply the general Cramer's rule for the derivation of the Löwdin formula and Thouless theorem. In the fifth section we derive the integral representation of Löwdin operator.

## 2. Generalization of Cramer's rule

Let E be a vector space of dimension (n) with basis $\vec{e}_1, \vec{e}_2, \ldots, \vec{e}_n$. $\vec{a}_1, \vec{a}_2, \ldots, \vec{a}_n$ Is a set of linearly independent vectors, belonging to E. $\vec{b}_1, \vec{b}_2, \ldots, \vec{b}_s$, $s \leq n$ is another set of linearly independent vectors, belonging to E. We denote by (A) the matrix formed by the components of the vectors $(\vec{a}_i)$ and $\det(A) = \det(\vec{a}_1, \vec{a}_2, \ldots, \vec{a}_n) = |(A)|$ is the determinant of the matrix (A).

**Theorem**: Consider the following systems

$$\sum_{j=1}^{n} \vec{a}_j x(k,j) = \vec{b}_k, \quad k = 1, 2, \ldots, s, \quad s \leq n \tag{2.1}$$

We find the determinant formed from det (A) by substituting the components of some vector $(\vec{a}_j)$ by the components of the vectors $\vec{b}_i$, $(1 \leq i \leq s) \leq n$ by the formula:

$$\det(a_1, \ldots, \vec{a}_{i_1}, \vec{b}_1, \ldots, \vec{a}_{i_s}, \vec{b}_s, \ldots, a_n) = \det(A) \begin{vmatrix} x(1, i_1) & \ldots & x(1, i_s) \\ \vdots & \vdots & \vdots \\ x(s, i_1) & \ldots & x(s, i_s) \end{vmatrix} \tag{2.2}$$

Prove: We will proceed by induction, s = 1 then 2, etc.

It is well known from the multilinear algebra [22] that the space $\overset{n}{\wedge} E$ has only one basic vector $\vec{e}_1 \wedge \vec{e}_2 \wedge \ldots \wedge \vec{e}_n$ and $\vec{a}_1 \wedge \vec{a}_2 \wedge \ldots \wedge \vec{a}_n = \det(A) \vec{e}_1 \wedge \vec{e}_2 \wedge \ldots \wedge \vec{e}$.

1-For s = 1 this is the case of Cramer's rule. We shall do a brief revision.

Multiply the two terms of the expression (2.1), the right by $\wedge \vec{a}_{i+1} \wedge \ldots \wedge \vec{a}_n$ and the left by $\vec{a}_1 \wedge \ldots \wedge \vec{a}_{i-1} \wedge$ we obtain:

$$\sum_{j=1}^{n} x(k,j) \vec{a}_1 \wedge \ldots \wedge \vec{a}_{i-1} \wedge \vec{a}_j \wedge \vec{a}_{i+1} \wedge \ldots \wedge \vec{a}_n = \vec{a}_1 \wedge \ldots \wedge \vec{a}_{i-1} \wedge \vec{b}_k \wedge \vec{a}_{i+1} \wedge \ldots \wedge \vec{a}_n$$

The summation in the first member is zero unless j = i, it follows that:
$$x(k,i)\vec{a}_1 \wedge \ldots \wedge \vec{a}_n = \vec{a}_1 \wedge \ldots \wedge \vec{a}_{i-1} \wedge \vec{b}_k \wedge \vec{a}_{i+1} \wedge \ldots \wedge \vec{a}_n \qquad (2.3)$$

then we deduce that
$$\det(A)x(k,i) = \det(\vec{a}_1,\ldots,\vec{a}_{i-1},\vec{b}_k,\vec{a}_{i+1},\ldots,\vec{a}_n) \qquad (2.4)$$

2-For s = 2, we multiply the two terms of (2.1), the right by
$$\wedge \vec{a}_{i+1} \wedge \ldots \wedge \vec{a}_{l-1} \wedge \vec{b}_r \wedge \ldots \wedge \vec{a}_n$$

And the left by $\wedge \vec{a}_{i+1} \wedge \ldots \wedge \vec{a}_{l-1} \wedge \vec{b}_r \wedge \ldots \wedge \vec{a}_n$,

We obtain:
$$\sum_{j=1}^{n} x(k,j)\vec{a}_1 \wedge \ldots \wedge \vec{a}_{i-1} \wedge \vec{a}_j \wedge \vec{a}_{i+1} \wedge \ldots \wedge \vec{a}_n =$$
$$\vec{a}_1 \wedge \ldots \wedge \vec{a}_{i-1} \wedge \vec{b}_k \wedge \vec{a}_{i+1} \wedge \ldots \wedge \vec{a}_{l-1} \wedge \vec{b}_r \wedge \ldots \wedge \vec{a}_n \qquad (2.6)$$

The first member is zero unless $j = i$ or $j = l$, it follows that
$$x(k,i)\det(\vec{a}_1,\ldots,\vec{a}_i,\ldots,\vec{a}_{l-1},\vec{b}_r,\vec{a}_{l+1},\ldots,\vec{a}_n) +$$
$$x(k,l)\det(\vec{a}_1,\ldots,\vec{a}_l,\vec{a}_{i+1},\ldots,\vec{a}_{l-1},\vec{b}_r,\vec{a}_{l+1},\ldots,\vec{a}_n) =$$
$$\det(\vec{a}_1,\ldots,\vec{a}_{i-1},\vec{b}_k,\vec{a}_{i+1},\ldots,\vec{a}_{l-1},\vec{b}_r,\vec{a}_{l+1}\ldots,\vec{a}_n) \qquad (2.7)$$

in the second term of the first member, we can interchange the order of vectors and we change the sign, by replacing the expressions of the first member using their value of (2.4) we get the expression.

$$\det(A)[x(k,i)x(r,l) - x(k,l)x(r,i)] = \det(A)\begin{vmatrix} x(k,i) & x(k,l) \\ x(r,i) & x(r,l) \end{vmatrix} \qquad (2.8)$$
$$= \det(\vec{a}_1,\ldots,\vec{a}_{i-1},\vec{b}_k,\vec{a}_{i+1},\ldots,\vec{a}_{l-1},\vec{b}_r,\vec{a}_{l+1},\ldots,\vec{a}_n)$$

3-We assume that (2.2) is true for s-1, we prove that is true for the case s.

Multiply the two terms of system (2.1), the left by
$$\vec{a}_1 \wedge \ldots \wedge \vec{a}_{i-1} \wedge \vec{b}_1 \wedge \ldots \wedge \vec{a}_{l-1} \wedge \vec{b}_{s-1}$$

and the right by $\wedge \vec{a}_{l+1} \wedge \ldots \wedge \vec{a}_n$, we obtain a similar expression of (2.7) and the summation is zero unless $j = i_1, i_2, \ldots, i_s$.

Using the result of the case s-1 and we note the minors by min, we find:
$$\det(A)[x(s,i_1)\min(x(s,i_1)) - x(s,i_2)\min(x(s,i_2)) + \ldots + x(s,i_s)\min(x(s,i_s))] =$$

$$= \det(A)\begin{vmatrix} x(1,i_1) & \ldots & x(1,i_s) \\ \vdots & \vdots & \vdots \\ x(s,i_1) & \ldots & x(i_s,i_s) \end{vmatrix} = \det(a_1,\ldots,\vec{a}_{i_1},\vec{b}_1,\ldots,\vec{a}_{i_s},\vec{b}_s,\ldots,a_n) \qquad (2.9)$$

### 3. The projection of the Hartree-Fock wave function

We present at first the basis of Hartree-Fock and then the calculation of the spectrum of rotations. For the calculation of the spectrum, we applied the projection of the Hartree-Fock wave function, and the application of Cramer's rule's generalization.

## 3.1 The Hartree-Fock basis

The variationnelle method leads to a hamiltonien called Hartree-Fock Hamiltonian whose eigenfunctions are the states of particles $\{|c_i\rangle\}$.

We denote the occupied states by $a_1, a_2, \ldots, a_n$ and $b_1, b_2, \ldots, b_i, \ldots$ the unoccupied states. In the second quantization formalism we write the wave functions of the system with the creation and destruction operators $\{a_i^+, a_j\}, \{b_k^+, b_l\}, 1 \leq i, j \leq n, 1 \leq k, l \leq \infty$ and we choose the wave function of Hartree-Fock as starting point.

$$|\Phi_{HF}\rangle = a_1^+ a_2^+ \ldots a_n^+ |0\rangle \qquad (3.1)$$

We note the states $\{|\Phi_i^j\rangle = b_j^+ a_i |\Phi_{HF}\rangle\}$ by particle-hole states $|1p-1h\rangle$ and the states $\{|\Phi_{ij}^{lm}\rangle = b_l^+ b_m^+ a_i a_j |\Phi_{HF}\rangle\}$ by the 2particles-2holes states $|2p-2h\rangle$, etc. All these states form a complete orthonormal basis which we call the Hartree-Fock basis.

## 3.2 The expression of the projection of the Hartree-Fock wave function

The spectrum of energy levels is given in the Peierls-Yoccoz theory [1-3] by

$$E_j = \frac{\int D_{(m,m)}^{*j}(\Omega) \langle \Phi_{HF} | HR(\Omega) | \Phi_{HF} \rangle d\Omega}{\int D_{(m,m)}^{*j}(\Omega) \langle \Phi_{HF} | R(\Omega) | \Phi_{HF} \rangle d\Omega} \qquad (3.2)$$

With H = T + V is the Hamiltonian, T is the kinetic energy and V is the potential energy. $\Omega = (\psi \beta \varphi)$ Is the solid angle and $D_{(m,m)}^j(\Omega) = e^{-im(\psi+\varphi)} d_{(m,m)}^j(\beta)$ is an element of rotations matrix.

In order to assure that the average value of H is minimal [], the variationnelle method imposes the condition:

$$\langle \Phi_{HF} | H | 1p-1h \rangle = \langle \Phi_{HF} | H b_j^+ a_i | \Phi_{HF} \rangle = 0, \forall i, j \qquad (3.3)$$

If we introduce the unitary operator of Hartree-Fock basis between H and R of the expression $\langle \Phi_{HF} | HR(\Omega) | \Phi_{HF} \rangle$ and taking into account the condition (3.3), we find in the case of axial symmetry:

$$E_j = E_{HF} + \frac{1}{4} \sum_{ij} \langle ij | \tilde{V} | kl \rangle \frac{\int d_{(m,m)}^{*j}(\beta) \langle \Phi_{HF} | a_i^+ a_j^+ b_l b_k e^{-i\beta J_y} | \Phi_{HF} \rangle \sin \beta d\beta}{\int d_{(m,m)}^{*j}(\beta) \langle \Phi_{HF} | e^{-i\beta J_y} | \Phi_{HF} \rangle \sin \beta d\beta} \qquad (3.4)$$

With $\langle ij | \tilde{V} | kl \rangle = \langle ij | V | kl \rangle - \langle ij | V | lk \rangle$

And $\langle ij | V | kl \rangle$ are the elements of the potential matrix.

We prove by simple calculation that [2]:

$$\langle \Phi_{HF} | e^{-i\beta J_y} | \Phi_{HF} \rangle = \det(\vec{a}_1, \vec{a}_2, \ldots, \vec{a}_n) \text{ With } a_{ij} = \langle a_i | e - i\beta J_y | a_j \rangle$$

And

$$\langle \Phi_{HF} | a_i^+ a_j^+ b_l b_k e^{-i\beta J_y} | \Phi_{HF} \rangle = \det(\vec{a}_1, \ldots, \vec{a}_{i-1}, \vec{b}_k, \vec{a}_{i+1}, \ldots, \vec{a}_{j-1}, \vec{b}_l, \vec{a}_{j+1}, \ldots, \vec{a}_n) \qquad (3.5)$$

With

$$(b_r)_m = \langle b_r | e^{-i\beta J_y} | a_m \rangle, 1 \le m \le n, \ r = k, l. \tag{3.6}$$

According to the preceding theorem, we deduce the final expression of energy.

$$E_j = E_{HF} + \frac{1}{4}\sum_{ij}\langle ij|\tilde{V}|kl\rangle \frac{\int d^{*j}_{(m,m)}(\beta)\langle \Phi_{HF}|e^{-i\beta J_y}|\Phi_{HF}\rangle \begin{vmatrix} x(k,i) & x(k,j) \\ x(l,i) & x(l,j) \end{vmatrix} \sin\beta d\beta}{\int d^{*j}_{(m,m)}(\beta)\langle \Phi_{HF}|e^{-i\beta J_y}|\Phi_{HF}\rangle \sin\beta d\beta} \tag{3.7}$$

We performed the calculations of $x(k,i)$ using the Gauss elimination method and the integration by Gauss method or Gauss-Legendre integration [11].

## 4. Applications of generalized Cramer's rule to the derivation of Löwdin formula and Thouless theorem

### 4.1 Generalization of Cramer's rule and Löwdin formula

We can extend the definition of variables $x(k,i)$ by

$$x(k,i) = \langle \Phi_{HF}|Rc_k^+c_i|\Phi_{HF}\rangle / \langle \Phi_{HF}|R|\Phi_{HF}\rangle \tag{4.1}$$

With $\qquad x(k,i) = 0, \ if \ i > n$

We find that $x(k,i) = \sum_{j=1}^{n}\langle c_k|R|a_j\rangle A_{ji}^{-1}$ and $A_{kj}^{-1}$ are the minor of the matrix (A).

We deduce that the one body potential may be written in the formalism of second quantization:

$$\langle \Phi_{HF}|TR|\Phi_{HF}\rangle = \sum \langle a_i|T|c_k\rangle \sum_k \langle c_k|R|a_j\rangle A_{ji}^{-1}\langle \Phi_{HF}|R|\Phi_{HF}\rangle \tag{4.2}$$

But $\sum_{k=1}^{\infty}|c_k\rangle\langle c_k|$ is the unitary operator in the space of one particle state.

Finally $\qquad \langle \Phi_{HF}|TR|\Phi_{HF}\rangle = \sum_{ij}\langle a_i|TR|a_j\rangle A_{ji}^{-1}\langle \Phi_{HF}|R|\Phi_{HF}\rangle \tag{4.3}$

Following the same method we find for the two body potential

$$\langle \Phi_{HF}|VR|\Phi_{HF}\rangle = \sum_{ijkl}\langle a_ia_j|\tilde{V}R|a_ka_l\rangle \left[\begin{vmatrix} A_{ki}^{-1} & A_{kj}^{-1} \\ A_{li}^{-1} & A_{lj}^{-1} \end{vmatrix}\right]\langle \Phi_{HF}|R|\Phi_{HF}\rangle \tag{4.4}$$

The inconvenient of Löwdin formula is the calculation of the elements $\{\langle a_ia_j|VR|a_ka_l\rangle\}$ that require long calculation.

## 4.2. Generalization of Cramer's rule and Thouless theorem

Let $|\Psi\rangle$ and $|\Phi\rangle$ be two wave functions such that $|\Psi\rangle = U|\Phi\rangle$, U is an invertible linear transformation and I is the unit operator of Hartree-Fock basis.

We have
$$U c_i^+ U^{-1} = \sum_j \langle c_j |U| c_i \rangle c_j^+ \qquad (4.5)$$

And
$$|\Psi\rangle = U|\Phi\rangle = IU|\Phi\rangle = \langle\Phi|U|\Phi\rangle|\Phi\rangle + \sum_{ik}\langle\Phi|a_i^+ b_k U|\Phi\rangle b_k^+ a_i |\Phi\rangle + \qquad (4.6)$$
$$\sum_{ijkl}\langle\Phi|a_i^+ a_j^+ b_k b_l U|\Phi\rangle b_l^+ b_j^+ a_j a_l |\Phi\rangle + \dots$$

Applying the theorem we get:
$$|\Psi\rangle = U|\Phi\rangle = IU|\Phi\rangle = \langle\Phi|U|\Phi\rangle|\Phi\rangle + \langle\Phi|U|\Phi\rangle\left[\sum_{ik} x(k,i) b_k^+ a_i |\Phi\rangle\right]$$
$$+ \langle\Phi|U|\Phi\rangle\left[\sum_{ijkl}\begin{vmatrix} x(l,i) & x(l,j) \\ x(k,i) & x(k,j) \end{vmatrix} b_l^+ b_k^+ a_j a_i |\Phi\rangle + \dots\right]$$
$$= \langle\Phi|U|\Phi\rangle\left[1 + (\sum_{ik} x(k,i) b_k^+ a_i) + \frac{1}{2!}(\sum_{ik} x(k,i) b_k^+ a_i)^2 + \dots\right]|\Phi\rangle \qquad (4.7)$$

This expression is written in the form

$$|\Psi\rangle = U|\Phi\rangle = \langle\Phi|U|\Phi\rangle \exp\left[\sum_{k,i} x(k,i) b_k^+ a_i\right]|\Phi\rangle \qquad (4.8)$$

In the particular case where $\langle\Phi|U|\Phi\rangle = 1$ we obtain the Thouless function [18].

## 5. Derivation of integral representation of Löwdin projection operator

The success of Hill-Wheeler [2] operator is due to the exponential form of the rotation operator so it is normal to find a similar expression for the infinitesimal operator. Thus, we find by a simple method the projections operators for the harmonic oscillator and angular momentum. Then we find the integral representation of Löwdin operator useful for the projection of Slater determinant.

We start from the fact that the states of classical groups [23-24] are ordered $|\varphi_1\rangle,\dots,$ $|\varphi_n\rangle,\dots$ and if we choose the projection $P_n |\varphi_n\rangle = |\varphi_n\rangle$ we can eliminate all the states before $|\varphi_n\rangle$, using the raising and lowering operators. We also find that these projectors have the form:

$$P_n(\gamma z)|\varphi_{n+i}\rangle = N \; _2F_1(,z)|\varphi_{n+i}\rangle, \quad i \geq 0 \qquad (5.1)$$

N, $\gamma_i$ are constants, z=1 is a root of the Gauss function, $_2F_1$ [18] and $P_n(\gamma) = P_n$.

## 5.1 projection operators of the harmonic oscillator

We want to determine the projector $p_n = |n\rangle\langle n|$ of the harmonic oscillator basis $\{|n\rangle\}$

with $\quad |\Phi\rangle = \sum_j C_j |j\rangle$ and $p_n |\Phi\rangle = C_n |n\rangle \quad$ (5.2)

$a^+, a$ are the creations and destructions operators of the harmonic oscillator defined by:

$$a|n\rangle = \sqrt{n}|n-1\rangle, \; a^+|n\rangle = \sqrt{(n+1)}|n+1\rangle \quad (5.3)$$

We find that

$$p_n |\Phi\rangle = \sum_{i=0} \gamma_i a^{+(n+i)} a^{(n+i)} |\Phi\rangle = \sum_j C_{n+j} \left( \sum_{i=1}^{j} \gamma_i \frac{(n+j)!}{(j-i)!} \right) |n+j\rangle \quad (5.3)$$

The identification of (5, 1) and (5.2) gives the triangular linear system of equations

$$p_j(\gamma) = \sum_{i=0}^{j} \gamma_i \frac{(n+j)!}{(j-i)!} = 0, \; \gamma_0 = 1, j, n \geq 0 \quad (5.4)$$

The solution of the triangular linear system $p_j(\gamma) = 0$ is:

$$\gamma_i = (-1)^n / n!. \quad (5.5)$$

Thus we get the projection operator $p_n = |n\rangle\langle n|$ of the harmonic oscillator:

$$p_n = \frac{1}{n!} \sum_{i \geq 0} \frac{(-1)^i}{i!} (a^+)^{n+i} (a)^{n+i} \quad (5.6)$$

We obtain the same expression of Schwinger [26] for n = 0 and $p_0 = |0\rangle\langle 0|$.

If we replace $\gamma_n$ by $\gamma_n z^n$ we find that the operator $P_n(\gamma z)$ has the properties

$$P_j(\gamma z)|n+j\rangle = \frac{(n+j)!}{n!} {}_2F_1(-j, \beta; \beta, z)|n+j\rangle \quad (5.7)$$

## 5.2 Integral representation of the infinitesimal projection operator

We derive first the expression of Löwdin projection operator and then its integral representation.

### 5.2.1 Expression of Löwdin projection operator $P_{jm}$

Let $|\Phi\rangle$ be a wave function of a system that has axial symmetry, admitting $m$ as eigenvalue of $J_z$. The development of $|\Phi\rangle$ on the basis $\{|jm\rangle\}$ is:

$$|\Phi\rangle = \sum_{j=|m|}^{\infty} C_j |jm\rangle \; . \quad (5.8)$$

It is well known from the theory of angular momentum that:

$$J_\pm |jm\rangle = \sqrt{(j \mp m)(j \pm m + 1)} |j(m \pm 1)\rangle, \quad (5.9)$$

And $\quad J_+ |jj\rangle = J_- |j-j\rangle = 0 \quad (5.10)$

Using the relation (5.9) and taking into account the relations (5.10) we find the system:

$$J_-^k J_+^k |\Phi\rangle = \sum_{l=j+k}^{\infty} C_j \frac{(l-m)!(l+m+k)!}{(l-m-k)!(l+m)!} |jm\rangle, \; k = 1, 2, etc. \quad (5.11)$$

We follow the same method as above for the derivation of the triangular linear system $P_r(\gamma z) = 0$. We get the solution:

$$\gamma_r = (-1)^r \frac{(2j+1)!}{(r)!(2j+r+1)!} \tag{5.12}$$

and the projection operator obtained by Löwdin [8] by a different method:

$$P_{jm} = \frac{(2j+1)(j+m)!}{(j-m)!} \sum_{r=0}^{\infty} (-1)^i \frac{J_-^{j-m+r} J_+^{j-m+r}}{r!(2j+r+1)!} \tag{5.13}$$

We also have $\quad P_r(\gamma z) = (j-m)!\, _2F_1(2j+(j-m)+1, -(j-m); 2j+2; z) \tag{5.14}$

For m = j, we obtain the so called extremal projection operators $P_j = P_{jj}$ [20].

It is important to emphasize that the operator $P_{00}$ in the Schwinger representation of angular momentum $\dfrac{(a_1^+)^{j+m}(a_2^+)^{j-m}}{\sqrt{(j+m)!(j-m)!}}|00\rangle$ is $|00\rangle\langle 00|$.

5.2.2 Integral representation of the infinitesimal projection operator

Using the expression of the integral [25]

$$\frac{(2j+1)(j-m)!}{(j+m)!} \int_0^1 dt \left(\frac{t^i}{(i!)^2}\right)(1-t)^{2m} P_{j-m}^{(0,2m)}(1-2t) = \frac{1}{r!(2j+r+1)!}, \tag{5.15}$$

With $i = j - m + r$ and $m \geq 0$.

Replacing this expression in (5, 14), we find by a simple calculation the integral representation of the infinitesimal projection operator.

$$P_{jm} = \int_0^1 P_{j-m}^{(0,2m)}(1-2t)(1-t)^{2m} e^{\bar{z}J_-} e^{zJ_+} dxdy,\ z = x + iy \tag{5.16}$$

With $z = \rho e^{-i\varphi}$, $t = \rho^2$